\documentclass[10pt]{article}
\usepackage[dvips]{graphicx}

\title{Square root relaxation: two possible mechanisms}
\author{Jacques Villain\\
D\'epartement de Recherche Fondamentale sur la Mati\`ere
Condens\'ee\\ D.S.M.,
Commissariat \`a l'Energie Atomique, 17 Avenue des Martyrs\\ F-38054 Grenoble
Cedex 9, France
}

\begin{document}
\maketitle

\begin{abstract}

Magnetic relaxation in large spin molecular paramagnets is often found to behave as
$\delta M(t)  \sim \sqrt{t}$ at short times $t$. This behaviour was 
explained by Prokofiev \& Stamp as arising from dipole interactions between molecular spins.
However, as observed by Miyashita \& Saito, the same behaviour can arise from 
a different mechanism which, in the present work, is related to hyperfine interactions.
 The Miyashita-Saito scheme is found to be possible 
at short times if the nuclear longitudinal
spin relaxation is very slow. In the case of moderately slow nuclear 
spin relaxation, the electronic magnetization variation $\delta M(t) $
is initially proportional to $t$, then to $\sqrt{t}$ and finally to $\exp(-t/\tau )$. 
This behaviour may be mostly expected in dilute systems. 

\end{abstract}

pacs 05.30.-d, 05.40.Fb, 75.50.Xx


Synthetic molecular 
nanomagnets\cite{Angew}  provide reproducible 
microscopic systems with a large magnetic moment which may have macroscopic properties.
The most widely studied materials are, in the usual terminology,
 Mn$_{12}$ac (with a relaxation time of 2 months for the magnetization at 2 K)
and Fe$_{8}$, whose faster relaxation allows easier experiments.

Magnetic relaxation of these materials at low temperature is a  challenging problem. 
The material is initially magnetized by a strong magnetic field $H^-$ 
in the easy magnetization direction $z$.
At $t=0$, the external field is suddenly given the value $ H_{ext} $,
also in the $z$ direction.  One measures the magnetization $ M_z (t)= M_z (0)+\delta  M_z (t) $. 
At low enough temperature, 
the following behaviour is observed for short times in  Fe$_{8}$  \cite{OSP,wern}
and Mn$_{12}$  \cite{kent,thom99}.
\begin{equation}
\delta M_z (t)= \delta M_z (\infty)  A  \sqrt{t}
\label{for1}
\end{equation}
where $A$ is a positive constant.
This square root behaviour is in contrast with 
usual relaxation which is exponential, and therefore linear for short times, 
$\delta M_z(t)=
\delta M_z(0) A' t$.

The square root behaviour (\ref{for1})
was predicted theoretically by Prokofiev \& Stamp \cite{prok} for the demagnetization
of a saturated sample. 
The molecule of interest can be modelled by a `molecular' spin ${\bf S}$,
of electronic nature,  whose modulus $s$ is large ($s=10$ for 
Mn$_{12}$ and Fe$_{8}$). This spin is
 subject to an anisotropy hamiltonian which, in Mn$_{12}$, may 
be written as a first approximation as

\begin{equation}
{\cal H}_0 = - KS_z^2
\label{form1}
\end{equation}
where the constant $K$ is positive.

Relaxation is slow because positive and negative values of $S_z$ 
form two different potential wells separated by a 
barrier of height $Ks^2$.
At temperatures and fields of interest ($k_BT \ll Ks^2$ and $g\mu_Bs H_{ext} \ll Ks^2$
where $\mu_B$ is the Bohr magneton and $g$ the Land\'e factor),  
thermal activation by phonon absorption is 
not possible and magnetic relaxation  
takes place by spin tunneling through the barrier under the effect of additional terms 
of the Hamiltonian which do not commute with $S_z$. Such a hamiltonian, adequate for Fe$_8$, is
${\cal H}={\cal H}_0+{\cal H}_1  = - KS_z^2 + g\mu_B H_z S_z +B S_x^2$,
where the magnetic field ${\bf H}$ has been introduced.

If ${\bf H}$ were constant, spin tunneling would be a periodic oscillation of $S_z$ 
between $-s$ and $s$, which  could be analyzed by
diagonalization of the hamiltonian. In the relaxation mechanism, the environment 
plays an essential part. It will be mimicked for each molecular spin by 
a time-dependent magnetic field $ H_z(t) =H_{ext} + \Delta H_z(t) $. The additional component
$\Delta H_z(t) $ is produced partly by the dipole interaction 
with the other molecular spins, and partly by the 
`hyperfine'  interaction with the nuclear spins. 
The transverse components $ \Delta H_x $ and $ \Delta H_y$, 
which slightly 
modulate the tunnel splitting $2\hbar \omega_T$, will be neglected. 
The $z$-component $ \Delta H_z$, though not larger, is very important.
Indeed, spin tunneling is only possible between 
two eigenstates of (\ref{form1}) which have nearly the same energy. This occurs
if, and only if the local longitudinal field
$H_z $ is close to 0 or to particular values (or `resonances') which depend on $K$. Attention will 
be focussed on one of these values which will be called $H_1$.
 Roughly speaking, the condition for tunneling  is that $ H_z $ satisfies
$ g\mu_B s |H_z-H_1 | < \hbar \omega_T$. 

Prokofiev \& Stamp \cite{prok} have shown that the  dipole interaction with the other molecular spins
leads to square root relaxation, formula (\ref{for1}).  In their theory, this
 is related to the $r^{-3}$ behaviour of
dipole interactions. It was later suggested by Miyashita \& Saito\cite{Miyashita} that  
square root relaxation
may follow from another, completely different mechanism which  may be equivalently termed
 `Wiener process', `random walk', or `diffusion'. 

In the present note, interactions between molecular spins are ignored. 
The square root law (\ref{for1}) is shown to arise from hyperfine interactions only
if 
nuclear spin relaxation is very slow.
If it is not so slow, our results are different from those of Miyashita \& Saito. 

A simplified model will be used. The assumptions are the following.

1) Interactions between molecular spins are neglected, $\Delta H_z(t) $ is 
the hyperfine field.

2) The hyperfine field $\Delta H_z(t)$ is a sum of 
independent random components $H_i^z(t)$ which can take $(2I_i+1)$ different values 
with a definite  probability per unit time
of jumping from one value to the other. The index $i$ labels the various nuclear spins. 

3) The number of nuclear spins interacting with a given molecular spin ${\bf S}$ is 
large.

4) Each molecular spin has a probability $\lambda $ per unit time  to relax
when the local
field $ H_z(t) =H_{ext} + \Delta H_z(t) $ acting on this spin is comprised
between $H_1 - \epsilon $ and $H_1 + \epsilon $, where 
$\epsilon \simeq \hbar \omega_T/(g\mu_Bs) $, otherwise 
there is no relaxation. 
The value of  $\lambda $ is expected to depend on the tunnel splitting, hyperfine field
 and nuclear spin dynamics. 

An extreme case is when a molecular spin has completely relaxed at time $t$
if, and only if, the local
field $ H_z(t) =H_{ext} + \Delta H_z(t) $ acting on this spin has been equal to
$H_1$ at some time $t'$ between 0 and $t$. This limiting case will be called 
`very slow nuclear relaxation'.

Assumption (1) is not claimed to be a good 
approximation, but a simplification consistent with our purpose, to investigate
$\sqrt{t}$ relaxation arising from hyperfine interactions only.

Assumption (2) implies that quantum coherence is lost after a short time. 
It is an oversimplification with respect to the real mechanism of nuclear relaxation.
In Mn$_{12}$ 
below about 0.1K, this mechanism seems to be complex, involving 
inhomogeneities of the crystal \cite{Morello}.

Assumption (3) is physically realistic since dipole interactions are long ranged.

Assumption (4) is appropriate when $H_1\neq 0$, so a spin can tunnel 
from the lowest state of its initial well to an excited state of the other well,
 where it deexcites with phonon emission. In the case $H_1=0$,
tunnelling takes place between the lowest states of each well
and phonons have no effect. Then assumption (4) must be reformulated.
This will be done in a separate section

\section{Analogy with a random walk}

According to assumption (4) the magnetization at time $t$  depends on the probability $p(h,t)$ 
that the hyperfine field $H^z(t') $ has taken the 
value $H_1$ for $0<t'<t$ if the initial field was  $ H^z(0) =h$. The field $H^z $
is  the sum of contributions $H_i^z $ of many nuclei. According
to assumption (2), these nuclei flip 
by random, uncorrelated jumps. Thus, they are similar to the steps of a random walker
In the simplest case,   the random walker 
has the same probability to go forward or backward. 
Then, the probability\footnote{To simplify the
language, the word `probability' will often be used instead of 
`density of probability'. The actual meaning  is clear from the formulae.} 
 $\rho(h,h',t)$ that a random walker is at $\Delta H_z=h'$ at time $t$ if he
started from $h$ at time 0 satisfies
 the diffusion equation
\begin{equation}
\frac{ \partial }{\partial t} \rho(h,h',t) =  D(h')  \frac{ \partial^2 }{\partial h'^2}  
\rho(h,h',t) 
\label{diffus}
\end{equation}
where $D $ is related to the relaxation time of nuclear spins. It will generally be assumed
to be independent of $h'$.
Then the solution of (\ref{diffus}) is 
\begin{equation}
 \rho(h,h',t) =  
 \frac{ 1 } { 2 \sqrt{\pi D t} } \exp \frac{-(h-h')^2}{4 D t}
\label{diffusol}
\end{equation}

\begin{figure}
\centering
  \includegraphics*[width=70mm, ]{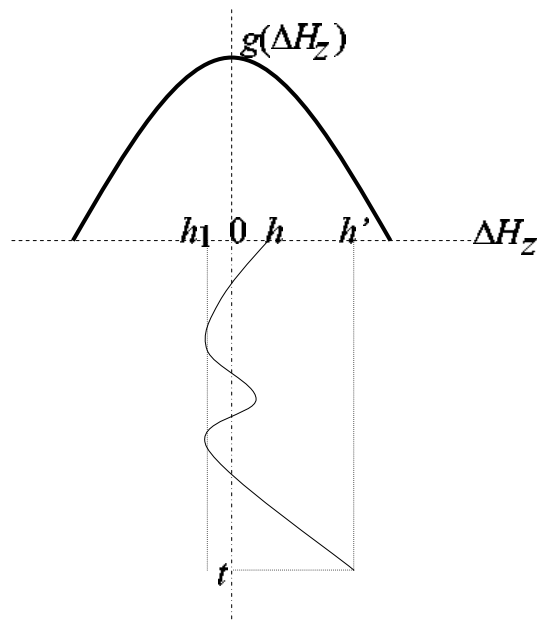}
\caption{ The thick curve shows the 
hyperfine field distribution $g(\Delta H_z ) $.
The thin curve shows a possible evolution with time $t$
of the hyperfine field $ \Delta H_z $
 on a particular molecular spin ${\bf S}$.}
\label{fig}
\end{figure}

Formulae (\ref{diffus}) and (\ref{diffusol}) are not valid for long times. 
Indeed the field distribution $g( \Delta H _z ) $
has a finite width $ \Delta H $ (Fig. \ref{fig}), so that the random walker
cannot reach fields higher than $\Delta H$. This can be accounted for by a force $f(h')$. 
The diffusion equation is thus replaced by the Fokker-Planck equation 

\begin{equation}
\frac{ \partial }{\partial t} \rho (h,h',t) =   \frac{ \partial }{\partial h'}
\left\{ 
D \left(  h'  \right) 
\left[ \frac{ \partial }{\partial h'}  \rho (h,h',t) 
- f(h') \rho (h,h',t) 
\right]
\right\}
\label{form166}
\end{equation}
which describes an Ornstein-Uhlenbeck process\cite{Miyashita}.
The force is easy to calculate by writing the equilibrium condition 
$   \rho (h,h',\infty) = g(h')$. It follows
\begin{equation}
f(h) =   \frac{ d}{dh} \ln g(h) 
\label{form167}
\end{equation}
If $g(h)$ is a gaussian, $g(h)=(2\pi)^{-1/2}\Delta H^{-1}\exp [-h^2/(2\Delta H^2)]$ 
then $f(h) = -h/\Delta H^2 $.

For short times, formulae (\ref{diffus}) and (\ref{diffusol}) 
are still correct in the presence of the force $f(h) $. 
Indeed, the displacement  of the random walker is the sum of 
a random part, of the order of $\sqrt{D t}$, and a drift part 
$D f(h)t \simeq -D ht/\Delta H^2 $. 
The former dominates the latter for short times. 

\section{Very slow nuclear spin relaxation}

The resonance condition $|H_z(t)-H_1|<\epsilon $ is satisfied during a time of the order of
$\epsilon ^2/D $. The flipping probability of a  molecular spin during this time is
$\lambda \epsilon ^2/D $. If this quantity is of order unity or larger, the nuclear relaxation
will be said to be `very slow'.
In that case, the proportion $n(t)$ of relaxed spins at time $t$ is

\begin{equation}
n(t)= \delta M_z(t) / \delta M_z(\infty) = 
\int_{-\infty}^\infty dh g(h) \int_{-\infty}^\infty dh' p_1(h_1,h,h',t)
\label{vercingetorix}
\end{equation}
where $p(h_1; h,h',t)$ is the probability that the random walker,
 who started from $h$
at time 0, is at $h'$ at time $t$ {\it and}
has reached $h_1=H_1 - H_{ext}$ at some time $t'$ between 0 and $t$. 
It is related to $ \rho( h,h',t) $ by

\begin{equation}
\left\{
\begin{array}{l}
p_1(h_1;h,h',t) = \rho( h,h',t) \;\;\;\;\hbox{if} \;\;\;\; (h-h_1)(h'-h_1)<0 \;\;\;\; \hbox{(a)} 
\\
\;\; \\
p_1(h_1;h,h',t) = \rho( 2h_1-h,h',t) \;\;\;\; \hbox{if} \;\;\;\; (h-h_1)(h'-h_1)>0 \;\;\;\; \hbox{(b)} 
\\
\end{array}
\right.
\label{archeopterix}
\end{equation}

Relation (\ref{archeopterix} a) is obvious and (\ref{archeopterix} b) can easily be derived, or 
found in textbooks\cite{feller}.

The integrand in (\ref{vercingetorix}) 
is very small unless $h-h_1$ and $h'-h_1$ are of the order of $\sqrt{D t}$
or smaller. In that domain the integrand is of the order of  $1/ \sqrt{D t}$. It follows that 
$\delta M(t) \sim \sqrt{D t}$ in agreement with (\ref{for1}).
The detailed calculation, based on (\ref{vercingetorix}), (\ref{archeopterix}) 
and  (\ref{diffusol}),   yields 

\begin{equation}
A =  8 g(h_1)   \sqrt{ D / \pi  }
  \int_0^{ \infty}  dx \int_0^{ \infty}  dy  \exp [-(x+y)^2] 
\label{formidable}
\end{equation}

There is no relaxation if 
$ |h_1| >  \Delta H $ (if resonances other than at $H_1$ are excluded). The maximum value 
of $A$ (obtained for $h_1=0$) is of the order of $1/\sqrt{\tau _H}$ where
\begin{equation}
\tau _H = \Delta H^2/D
 \label{long}
\end{equation} 
is the longitudinal nuclear spin relaxation time. The notation $T_1$ has been avoided because
it usually designates the spin-lattice relaxation time, related to spin-phonon interactions,
which is extremely long at low temperature. The notation $T_2$ has also been avoided because
it usually designates the transverse nuclear spin relaxation time.


\section{Moderately slow nuclear spin relaxation}

The nuclear spin relaxation will be called moderately slow when  the relaxation of a
molecular,  electronic spin requires that $(H_1-H_z)$ vanishes several times, but is
already almost complete (at resonance, i.e. $h_1=0$) at $t=\tau _H$.

A qualitative description will be given in the case $ h_1 =H_1-H_{ext}=0$.
 As seen from (\ref{diffusol}), the random walker
explores in the time $t$ a field interval 
$ \delta h(t) \simeq \sqrt{D t} $.
 The proportion of field values which have been  explored at time $t$ is thus

\begin{equation}
p(t) \simeq \delta h(t) /\Delta H \;\;\;\; .
 \label{modera1}
\end{equation}  

This quantity $p(t)$ is also the proportion of molecular
spins which have a chance to relax during the time $t$.  
 For such a spin,  tunneling is allowed 
during a time
\begin{equation}
 t_{eff}  \simeq   \epsilon t / \delta h(t) \simeq  \epsilon  \sqrt{t/D } \;\;\;\;.
  \label{modera2}
\end{equation} 

For a spin whose local field  lies in the explored region $\delta h(t) $,
the relative magnetization change, for short times,  is $ \lambda t_{eff} $, 
and this evaluation is correct if 
$  \lambda t_{eff} <1$. 
The total relative magnetization change is for such times 
$$
 \delta M_z(t) /\delta M_z(\infty) = p(t)  \lambda t_{eff} 
$$
or, according to (\ref{modera1}) and (\ref{modera2}),
 \begin{equation}
 \delta M_z(t) /\delta M_z(\infty)  \simeq  \lambda \epsilon t/\Delta H\;\;.
 \label{modera}
\end{equation} 

This formula is different from that written by Miyashita \& Saito, who assume $h=h_1$
at $t=0$, so that $p(t) $ is replaced by 1
in (\ref{modera}). Therefore, they predict  
$ \delta M_z(t) \sim \sqrt{t}$
at short time instead of  $\delta M_z(t) \sim {t}$ as obtained in ( \ref{modera}).
The quantity they calculate is actually $ \lambda t_{eff} \simeq  \lambda  \epsilon  \sqrt{t/D } $, 
which is given by their formula (3.9) as $\alpha \hbar^2 \omega _T^2 \sqrt{t/D}$ 
where $\alpha $ is a constant. Since $\epsilon \simeq \hbar \omega_T/(g\mu_Bs) $, 
identification of the two results yields
\begin{equation}
 \lambda  = \omega _T \tau _X
 \label{moderator}
\end{equation} 
where $\tau _X$ is a constant time (related to $\alpha $). Thus, Miyashita \& Saito
find that $ \lambda $ is independent of $D$ and $\Delta H$. 
This surprising result is based on the assumption
of a fairly {\it fast} nuclear spin
relaxation, and on the statement that ``the velocity of the field is proportional to  $\sqrt{D}$''.
This point will not be discussed here.

When $t$ becomes so large that $  \lambda t_{eff} >1$,
the depolarization for  an `explored'
field  value is almost total and the average relative demagnetisation is 
$ \delta M_z(t) /\delta M_z(\infty) = p(t) \simeq \sqrt{D t} / \Delta H $ $= \sqrt{t\tau_H}$
as in the case of very slow nuclear relaxation. This result holds if $ |h_1 |$
is appreciably smaller than $\Delta H $. 
When $ |h_1 |$ approaches $\Delta H $, 
the rate of change of the magnetisation can easily be shown to decrease to 0.


\section{Long times}

For times $t \gg \tau _H$, the random walker has lost the memory of its initial position $h$.
Thus, to quote Miyashita \& Saito, ``the number of crossings'' [of $H_z(t) $ with $H_1$
per unit time] `` is  ... constant in time, which causes a constant rate relaxation, i.e.,
the exponential relaxation.'' Two remarks should be added. First, the electronic 
spin relaxation time $\tau $
obviously depends on $h_1=H_1 - H_{ext}$.  Miyashita \& Saito made detailed
numerical studies, but only  in the case 
of an external field tuned at resonance, $h_1=0$. When $h_1\neq 0$ becomes larger than the  
width $\Delta H$, $\tau $ obviously goes to $\infty$.
The second remark is that, in the
case of very slow or moderately slow nuclear spin resonance, which is addressed in the present note, 
the electronic spin relaxation is already very strong at $t = \tau _H $, i.e. 
$\delta M( \tau _H )/ \delta  M(\infty)$ is of order unity. This implies that 
the exponentially relaxing part is  small.


\section{ The case $H_1=0$}
In most of the experiments done so far, $H_1=0$, i.e.
 tunnelling takes place between the lowest state of each well.  This case 
is peculiar. 1) The main feature is that (if phonons are completely ignored) the 
system of electronic and nuclear spins ignore the temperature $T$ of the crystal, and 
relax to 
an equilibrium state which generally corresponds to another temperature $T_{eff}$. 
Indeed, only a part of the Zeeman energy of the molecular spins can be transferred 
to nuclear degrees of freedom. Only for $H_{ext}=0$,  when the Zeeman energy vanishes, 
$T_{eff}=T$.
2) If nuclear spin relaxation is not very slow, the above treatment is still acceptable for 
short times, because spin reversal and relaxation are nearly the same thing.
3) Let the case of very  slow nuclear spin relaxation
be discussed. Then, when the hyperfine field $\Delta H_z$ crosses the value $-H_{ext}$,
$S_z$ follows the field adiabatically and changes sign. 
Thus, if $S_z(0)=s $ then $S_z(t)=s $ 
if $(h-h_1)$ and $(h'-h_1)$ have the same sign, while  $S_z(t) =-s$ if they
have a different sign. Thus, the spin never loses the memory of its initial state, 
which is quite unusual in a relaxation processs.
For short times, formula (\ref{for1}) can then be derived from an argument similar, 
but not identical to the above one, and $A$ turns out to be given by (\ref{formidable})
again. For long times, the 
 hyperfine field distribution is probably affected by relaxation and  depends on time.


\section{ Validity of the very slow nuclear relaxation scheme}

For a spin able to tunnel between two localized states $ |-\rangle $  
and $|+\rangle$, the wave function $x(t) |+\rangle + y(t) |-\rangle $ satisfies

\begin{equation}
\dot{x}(t)
= \frac{1}{i\hbar} x(t) E^{(+)}(t)
-i\omega_T y(t)
\;\;\;\;;\;\;\;\; 
 \dot{y}(t)
=-i\omega_T x(t) + \frac{1}{i\hbar} y(t) E^{(-)}(t)
\\
\label{eq4c22}
\end{equation}
where the unperturbed energies $ E^{(\pm)}(t)$ satisfy 
$ E^{(-)}(t) -  E^{(+)}(t)=2g\mu_B [H_z(t)-H_1] s$.

Equation simplifies if one introduces the notations
$u(t)=(1/\hbar) \int_{t_0}^{t} dt' E^{(+)}(t')\;$, 
$w(t)= \frac{1}{\hbar} \int_{t_0}^{t} dt'  E^{(-)}(t')\;$, 
$x(t) = \exp[-iu(t)] X(t)\; $,
and $y(t) = \exp[-iw(t)] Y(t) $. Moreover, the initial condition $X(0)=1$ will be assumed, 
and the time
will be assumed so short that $X(t)$ may be approximated by  $X(0)=1$. Then
(\ref{eq4c22}) yields $\dot{Y}(t) = -i\omega_T exp[-i U(t)] $
where

\begin{equation}
U(t) = (2g\mu_Bs/\hbar) \int_{t_0}^{t} dt' [H_z(t')-H_1] 
\label{nucspin107}
\end{equation}

Let the initial value of the local field be  tuned so as to allow tunnelling. If 
the tunnelling window remains open until a time of the order of the tunnel period $1/\omega _T$,
nuclear relaxation is very slow in the sense defined above.
The condition for very slow relaxation is thus $U(1/\omega _T)<1$.
 According to (\ref{nucspin107}),
$U(t) $ can be roughly evaluated as the product of $(2g\mu_Bs/\hbar) $ by 
$ \sqrt{D t} = \Delta H \sqrt{ t/ \tau _H } $
and a factor $t$ because of the integration. The condition for very slow relaxation reads 
\begin{equation}
 \frac{2g\mu_Bs \Delta H }{\hbar \omega _T \sqrt{ \omega _T  \tau _H }} < 1
\label{slow}
\end{equation}

The hyperfine width $\Delta H $ is never smaller than 0.001 Tesla. Thus, (\ref{slow})
requires large values of $\tau _H $ and $\omega _T  $.
For instance, a nuclear relaxation time $\tau _H $=1 s implies $ \omega _T \geq 10^{6}$ s$^{-1}$. 

An additional condition is $ \omega _T \tau _H >1$, but condition (\ref{slow}) is probably stronger.

 Usual nuclear spin-lattice 
relaxation is expected to be extremely slow at low temperature. However, 
measurements  by Morello et al. \cite{Morello} in Mn$_{12}$ac reveal  that $\tau _H$
does not increase beyond about 0.01 second.  The zero field tunnel splitting of the ground doublet
does not seem to satisfy the requirement $\omega _T \tau _H>1$ (Barbara, private communication).  
Thus, nuclear spin relaxation is not very slow and therefore the hyperfine field is not
expected to give rise to $\sqrt{t}$ relaxation in zero external field
in Mn$_{12}$ac, but rather to exponential relaxation. 

It is of interest to recall that square root relaxation has been observed in Fe$_8$ on times
of the order of a minute. The nuclear spin lattice relaxation times $T_1$ which have been reported
(for instance when observing negative spin temperatures\cite{goldman}) are of the same order 
of magnitude,
but it is not clear whether so long  nuclear longitudinal relaxation times can be reached in
molecular nanomagnets.


\section{Real systems: effect of dipole interactions}

In real systems, dipole interactions between molecular spins are present and contribute
to the local field by an amount which will be called `dipole field'. This contribution 
adds to the hyperfine field addressed above. Total magnetic relaxation
requires that the dipole field, as well as the hyperfine field, explores the whole  allowed region.
In Fe$_8$, the dipole field created by electronic spins  
is about 10 times as large as the hyperfine field, and therefore 90\% 
of the relaxation cannot be explained by the Miyashita-Saito mechanism described above, 
but only by the Prokofiev-Stamp mechanism.


\section{Conclusion}

In the present work it is shown that the $\sqrt{t}$ behaviour can in principle arise 
from hyperfine interactions alone. 
This can happen for any value of the external field and for any initial state provided it 
is not the equilibrium state. 
However, very slow nuclear spin relaxation is necessary. In the case of moderately slow nuclear spin relaxation, the decay is found to be linear 
at short times, in contrast with the statements of Miyashita \& Saito. This discrepancy 
occurs because  Miyashita \& Saito assumed the initial condition $h=h_1$ which  
is not fulfilled by nuclear spins.
After some time the decay  crosses over to the $\sqrt{t}$ behaviour and finally becomes 
exponential at long times. The constant $A$ of formula  (\ref{for1}), and the relaxation 
rate $1/\tau $ which describes the long time behaviour, go to 0 when the tuning parameter
$(H_{ext}-H_1)$ becomes large with respect to the hyperfine width $\Delta H$.   

When $H_1=0$ and $H_{ext}\simeq 0$, spin relaxation has remarkable features when the
spin-lattice relaxation time is very long. The spin temperature can then be quite different 
from the lattice temperature.

The present work contains several shortcomings. The possibility of simultaneous reversal of the 
electronic spin and a few nuclear spins has been disregarded. Such processes have been treated 
by Prokofiev \& Stamp\cite{prokostamp} in the case of superparamagnetic grains. They might be
less crucial for molecular nanomagnets since each molecular spin interacts in a
significant way with a restricted number of nuclear spins. 
On the other hand, the possible dependence of $D$ and $\lambda $ 
with respect to the local field has also been
ignored (as in the work of Miyashita \& Saito). Such a dependence would modify the value   
(\ref{formidable}) of the coefficient $A$.

The present theory may be relevant, for instance, in two cases. i) Mn$_{12}$ac 
if a transverse field is applied in order to increase $\omega _T$
and to fulfill the condition (\ref{slow}) of very slow nuclear spin relaxation. 
ii) Diluted samples of Fe$_{8}$, where the dipole interaction between molecular spins   is
smaller than the hyperfine field.

\vskip1cm

I acknowledge precious informations from Bernard Barbara,
Claude Berthier, Eugene Chudnovsky, Bernard Derrida, 
Julio Fern\'andez, Maurice Goldman and Wolfgang Wernsdorfer

\end{document}